\documentclass[a4paper,english]{iopart}
\usepackage[T1]{fontenc}
\usepackage[latin9]{inputenc}
\usepackage{amstext}
\usepackage[pdftex]{graphicx}

\makeatletter

\providecommand{\tabularnewline}{\\}

\usepackage{iopams}
\usepackage{setstack}

\@ifundefined{showcaptionsetup}{}{%
 \PassOptionsToPackage{caption=false}{subfig}}
\usepackage{subfig}
\makeatother

\usepackage{babel}

\begin{document}

\title{Effects of Disorder in FeSe : An \emph{Ab Initio} Study}

\author{Prabhakar P. Singh}

\ead{ppsingh@phy.iitb.ac.in}

\address{Department of Physics, Indian Institute of Technology Bombay, Powai,
Mumbai- 400076, India }
\begin{abstract}
Using the coherent-potential approximation, we have studied the effects
of excess Fe, Se-deficiency, and substitutions of S, Te on Se sub-lattice
and Co, Ni and Cu on Fe sub-lattice in FeSe. Our results show that
(i) a small amount of excess Fe substantially disorders the Fe-derived
bands while Se-deficiency affects mainly the Se-derived bands, (ii)
the substitution of S or Te enhances the possibility of Fermi surface
nesting, specially in FeSe$_{0.5}$Te$_{0.5}$, in spite of disordering
the Se-derived bands, (iii) the electron doping through Co, Ni or
Cu disorders the system and pushes down the Fe-derived bands, thereby
destroying the possibility of Fermi surface nesting. A comparison
of these results with the rigid-band, virtual-crystal and supercell
approximations reveals the importance of describing disorder  with
the coherent-potential approximation. 
\end{abstract}



\submitto{JPCM}

\maketitle

\section{Introduction}

The superconductivity in iron pnictides \cite{Kamihara,ZARen} and
chalcogenides \cite{FCHsu}, with the interactions resulting from
charge and spin degrees of freedom of electrons delicately balanced
\cite{AJDrew,WBao,MJHan}, provide a unique opportunity to unravel
the mystery of unconventional superconductivity \cite{NFBerk,PMon,TMoriya,TMoriyabook}.
From the experimental characterizations \cite{KIshida} of the superconducting
state of the iron pnictides and chalcogenides as well as theoretical
calculations \cite{IMazin1}, it seems clear that the superconductivity
in these alloys is not mediated by the phonons \cite{ASubedi}. Therefore,
attempts are being made to understand the superconductivity in iron
pnictides and chalcogenides in terms of spin-fluctuations and related
theories \cite{IMazin2}. 

In spin-fluctuation theories \cite{TMoriya,TMoriyabook,IMazin2},
in addition to the proximity of the system to a magnetic instability,
its Fermi surface (FS) and, in particular, the FS nesting plays a
crucial role in enhancing the magnetic interactions. In turn, by tuning
the magnetic properties and the FS of the system by chemical substitutions,
electron or hole doping, one can bring out the details of the superconducting
properties of the system, and thereby close in on the exact nature
of interaction responsible for superconductivity. Out of all the recently
discovered, superconducting iron pnictides and chalcogenides, FeSe
and its alloys with one of the simplest crystal structure are well-suited
for such a study. 

In FeSe, the superconducting transition temperature $T_{c}$ of $\simeq8$
K \cite{FCHsu}, increases up to $15$ K with S and Te substitutions
\cite{YMizuguchi2} but decreases rapidly with electron doping using
Co, Ni and Cu substitutions \cite{YMizuguchi2,TMM2,AJWilliams}. The
presence of excess Fe as well as the Se deficiency in FeSe is known
to affect its superconducting properties \cite{TMMcQueen}. In addition,
with increase in pressure up to $8.9$ GPa, the $T_{c}$ of FeSe increases
to $36.7$ K \cite{YMizuguchi,SMargadonna,SMedvedev}. 

Previous theoretical attempts \cite{ANYaresko,KWLee,LZhang,GXu,SLYu}
at understanding the changes in the normal state electronic properties,
as a prelude to understanding the changes in their superconducting
properties, of several of the pnictides and chalcogenides upon alloying
have used either virtual-crystal or the supercell or a series of ordered
alloys or parametrized model Hamiltonian approach. The virtual-crystal
and the supercell approaches are known to be quite inadequate to describe
the effects of disorder in metallic alloys, especially in $d$-band
metals \cite{faulkner}. In addition, an \emph{ab initio} study of
the effects of alloying in FeSe has been lacking so far. 

In an attempt to understand the changes in the normal state electronic
properties of FeSe upon alloying, which may, in turn, provide some
understanding of the changes in its superconducting properties, we
have used Korringa-Kohn-Rostoker coherent-potential approximation
method \cite{faulkner} in the atomic-sphere approximation (KKR-ASA
CPA) \cite{PPSingh1} to carry out charge, self-consistent electronic
structure calculations of FeSe and its alloys with S, Te, Co, Ni and
Cu, including the presence of excess Fe and the deficiency of Se.
We have also carried out a detailed comparison of these results with
the rigid-band, virtual-crystal and supercell approximations. 

Based on our calculations, we find that (i) a small amount of excess
Fe substantially disorders the Fe-derived bands near Fermi energy
($E_{F}$) while Se-deficiency affects mainly the Se-derived bands
away from $E_{F}$, (ii) the substitution of S or Te on Se sub-lattice
enhances the possibility of FS nesting in spite of disordering the
Se-derived bands, specially in FeSe$_{0.5}$Te$_{0.5}$ alloy , (iii)
the substitution of Co, Ni or Cu on the Fe sub-lattice disorders the
system and pushes down the Fe-derived bands, thereby destroying the
possibility of FS nesting. We also find that for describing substitutional
disorder the coherent-potential approximation is more reliable than
the rigid-band, virtual-crystal or supercell approximation.

\section{Computational Method\label{sec:Computational-Method}}

We have studied Fe$_{1+\delta}$Se with $\delta=0.0,0.01,0.03,0.06$,
FeSe$_{1-x}$ with $x=0.01,0.03,0.06,0.12,0.18$, FeSe$_{1-y}$Te$_{y}$
with $y=0.25,0.5$, FeSe$_{0.9}$S$_{0.1}$, Fe$_{0.9}$Co$_{0.1}$Se,
Fe$_{0.9}$Ni$_{0.1}$Se and Fe$_{0.91}$Cu$_{0.1}$Se in the tetragonal
(P4/nmm) crystal structure, while Fe$_{1.01}$Se has been studied
in the orthorhombic structure as well. The substitution of 10\% Cu
on Fe sub-lattice in Fe$_{1.01}$Se results in the composition Fe$_{0.91}$Cu$_{0.10}$Se,
used in the cited experiment \cite{AJWilliams} and hence, in our
calculations. To reduce the errors due to the ASA, we have introduced
four empty spheres in the unit cell containing two Fe and two Se atoms.
The atomic as well as the empty-sphere positions in the unit cell
are: Fe $(3/4,1/4,0;\,2a)$, Se $(1/4,1/4,z_{Se};\,2c)$, E1 $(1/4,1/4,z=-z_{Se};\,2c)$
and E2 $(3/4,1/4,0.5;\,2c)$, where E1 and E2 denote the two empty-sphere
sub-lattices. The E1 empty sphere layer, in the same plane as the
Se-layer, were used to accommodate the excess Fe in the calculations
of Fe$_{1+\delta}$Se and Fe$_{0.91}$Cu$_{0.1}$Se. To see the effects
of incorporating the excess Fe in other interstitial regions, we have
also used the E2 empty-sphere site for the excess Fe in Fe$_{1.06}$Se
and sites just below the Se-plane.

To model the effects of disorder, we have used the CPA \cite{soven,faulkner}
rather than a rigid-band, virtual-crystal or supercell method because
CPA has been found to reliably describe the effects of disorder in
metallic alloys \cite{PPSingh1,faulkner}. We used Barth and Hedin
\cite{BarthHedin} exchange-correlation potential. The Brillouin zone
(BZ) integration during self-consistency was carried out using a grid
of 24x24x20 points in the BZ. The density of states (DOS) was calculated
with a grid of 28x28x24 points in the BZ except for the pure FeSe
and FeTe, where a grid of 36x36x32 points was used. For both DOS and
spectral function calculations, we have added a small imaginary component
of $1$ mRy (1.5 mRy for FeSe and FeTe) to the energy. In the following
figures the Se-derived $s$-band is not shown. In our calculations,
the lattice parameters $a$ and $c$ were taken from experiments \cite{YMizuguchi2,TMMcQueen,TMM2,AJWilliams}
while the theoretically relaxed $z$-values were taken from Ref. \cite{ASubedi}.
For some of the intermediate alloys, we used the concentration-weighted
average of the $z$-values. The lattice parameters of FeSe, FeTe and
their alloys, used in the present calculations, are listed in Table
\ref{tab:lattice}. 

We have analyzed our results using the Bloch spectral function \cite{faulkner}
$A(\mathbf{k},E)$, defined by \begin{equation}
A(\mathbf{k},E)=-\frac{1}{\pi}\Im G(\mathbf{k},E),\label{eq:spf}\end{equation}
where $G(\mathbf{k},E)$ is the $\mathbf{k}$-space coherent-potential
Green's function, and $\mathbf{k}$ and $E$ represent the wave vector
and the energy, respectively, of the electron. The band structure
along BZ symmetry directions was calculated by evaluating Eq. (\ref{eq:spf})
for the given $\mathbf{k}$ points and the energy $E$ in the given
range. The Fermi surface in a given $\mathbf{k}$-space plane was
mapped by evaluating Eq. (\ref{eq:spf}) over a two-dimensional grid
of 151$\times$151 \textbf{$\mathbf{k}$} points in the plane at the
Fermi energy $E=E_{F}.$ The peaks in the spectral function $A(\mathbf{k},E_{F})$
form the Fermi surface of the alloy. All the plots of the band structures
and the Fermi surfaces were carried out with the same value for the
colormap in the range of $0$ to 30 (15 for the Fermi surface) with
the minimum and maximum value represented by blue and red, and the
intermediate values assigned colors varying from blue, light blue,
green, yellow, orange and then red. Thus, a movement away from red
and towards blue in the band structure and Fermi surface plots indicate
reduction in the peak value of $A(\mathbf{k},E)$.

\begin{table}
\caption{The experimental lattice parameters $a$ and $c$ of FeSe, FeTe and
their alloys used in the present calculations. \label{tab:lattice} }

\begin{tabular}{|c|c|c||c|c|c|}
\hline 
Alloy & $a$ ($\textrm{\AA)}$ & $c$ $\textrm{(\AA)}$ & Alloy & $a$ ($\textrm{\AA)}$ & $c$ $\textrm{(\AA)}$\tabularnewline
\hline
\hline 
FeSe & 3.765 & 5.518 & Fe$_{0.90}$Co$_{0.10}$Se & 3.7637 & 5.5043\tabularnewline
\hline 
Fe$_{1.01}$Se, FeSe$_{0.99}$ & 3.7734 & 5.5258 & Fe$_{0.90}$Ni$_{0.10}$Se & 3.7713 & 5.503\tabularnewline
\hline 
Fe$_{1.03}$Se, FeSe$_{0.97}$ & 3.7787 & 5.5208 & Fe$_{0.91}$Cu$_{0.10}$Se & 3.807 & 5.495\tabularnewline
\hline 
Fe$_{1.06}$Se, FeSe$_{0.94}$ & 3.7747 & 5.5229 & FeSe$_{0.90}$S$_{0.10}$ & 3.763 & 5.503\tabularnewline
\hline 
Fe$_{1.125}$Se & 3.7747 & 5.5229 & FeSe$_{0.75}$Te$_{0.25}$ & 3.7989 & 5.9685\tabularnewline
\hline 
FeSe$_{0.88}$, FeSe$_{0.875}$ & 3.7693 & 5.4861 & FeSe$_{0.50}$Te$_{0.50}$ & 3.7909 & 5.9570\tabularnewline
\hline 
FeSe$_{0.82}$ & 3.7676 & 5.4847 & FeTe & 3.8215 & 6.2695\tabularnewline
\hline
\end{tabular}
\end{table}

The formation energy, $E_{form},$ of excess Fe in Fe$_{1+\delta}$Se
and Se vacancy in FeSe$_{1-x}$ alloys has been calculated as \begin{equation}
E_{form}^{e}=E_{Fe_{1+\delta}Se}-E_{FeSe}-\delta E_{Fe}\label{eq:formation1}\end{equation}

\begin{equation}
E_{form}^{v}=E_{FeSe_{1-x}}-E_{FeSe}+xE_{Se}\label{eq:formation2}\end{equation}
where the superscripts $e$ and $v$ correspond to the formation energy
of FeSe with excess Fe or with Se vacancy, respectively. The subscripts
on the right hand side of Eqs. (\ref{eq:formation1}) and (\ref{eq:formation2})
denote the calculated total energies of the corresponding alloys except
for $E_{Fe}$ and $E_{Se}$ which represent the corresponding free
atom energy. 

The rigid-band calculations were carried out by first determining
the amount of shift in Fermi energy required to either accommodate
or remove a given number of electrons from the self-consistent potential
of ordered FeSe. The Fermi surface in the rigid-band was then evaluated
with the KKR-ASA CPA method at the shifted Fermi energy and using
the self-consistent potential of ordered FeSe. 

For the virtual-crystal approximation \cite{JKubler}, we replaced
the disordered sub-lattice by an ordered sub-lattice containing a
$virtual$ atom, and then carried out a charge, self-consistent calculation
for the alloy with the virtual atom using the KKR-ASA CPA method.
The virtual atom on a sub-lattice was constructed by adding the concentration-weighted
average of the valence electrons of the respective atoms on the sub-lattice
to the core electrons of the atom having higher concentration. For
example, the atomic number of the virtual atom on the Fe sub-lattice
in Fe$_{0.9}$Ni$_{0.1}$Se alloy is $26.2.$

\begin{figure}
\subfloat[]{\includegraphics[bb=14bp 14bp 345bp 504bp,clip,scale=0.22]{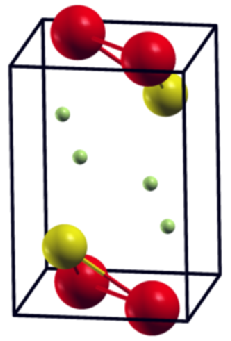}}\subfloat[]{\includegraphics[scale=0.25]{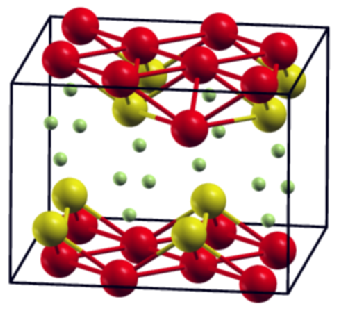}}\subfloat[]{\includegraphics[scale=0.25]{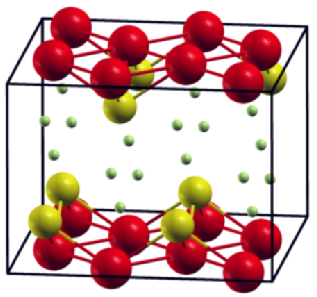}}

\caption{(Color online) The unit cell of (a) tetragonal FeSe and the 2$\times$2$\times$1
supercells used for (b) Fe$_{9}$Se$_{8}$ and (c) Fe$_{8}$Se$_{7}$
calculations. The large spheres represent Fe (red) and Se atoms (yellow),
and the smaller spheres represent empty-spheres (grey). The excess
Fe in Fe$_{9}$Se$_{8}$ is placed at an empty-sphere site while vacancy
in Fe$_{8}$Se$_{7}$ is formed by replacing a Se atom by an empty-sphere.
\label{fig:str-2-9-7}}

\end{figure}

In the supercell approximation for excess Fe in Fe$_{1.125}$Se and
Se vacancy in FeSe$_{0.875}$, we used a $2a\times2a\times c$ supercell
of FeSe containing 32 sites (16 atomic and 16 empty-sphere sites)
as shown in Fig. \ref{fig:str-2-9-7}. The excess Fe in Fe$_{9}$Se$_{8}$
is placed at an empty-sphere site while vacancy in Fe$_{8}$Se$_{7}$
is formed by replacing a Se atom by an empty-sphere. The charge self-consistent
calculations for the supercells were carried out using the KKR-ASA
CPA method. The Brillouin zone (BZ) integration during self-consistency
was carried out using a grid of 12x12x20 points in the BZ. The density
of states (DOS) was calculated with a grid of 14x14x24 points in the
BZ. For DOS and spectral function calculations, we have added a small
imaginary component of $2$ mRy and 1 mRy, respectively, to the energy.
We have also checked our results of the supercell calculations with
the LMTO-ASA method.

\section{Ordered $FeSe$ and $FeTe$ in the atomic- sphere approximation\label{sec:fese-fete}}

\begin{figure}
\subfloat[]{\includegraphics[scale=0.5]{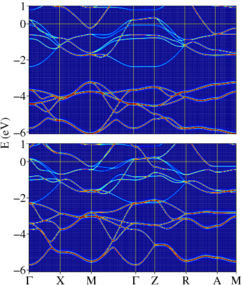}

}\subfloat[]{\includegraphics[scale=0.39]{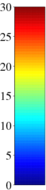}}\caption{(Color online) (a) The band structure of tetragonal FeSe (top) and
FeTe (bottom) along the BZ symmetry directions using experimental
lattice parameters and theoretically relaxed $z_{Se}$. The Fermi
energy is indicated by the horizontal line at $0$ eV. (b) The colormap
used for all the band structure and the Fermi surface plots as described
in the text. \label{fig:bs-fese-fete}}

\end{figure}

\begin{figure}
\includegraphics[clip,scale=0.35]{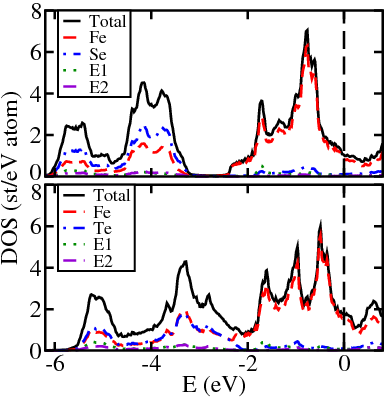}\caption{(Color online) The total (black, solid) and the sub-lattice-resolved
DOS of tetragonal FeSe (top) and FeTe (bottom) obtained using experimental
lattice parameters and theoretically relaxed $z_{Se}$. The contributions
from the Fe (red, dash) and the Se/Te (blue, dot-dash) sub-lattices
as well as from the empty-sphere sub-lattices E1 (green, dot) and
E2 (violet, double-dash dot) are shown. The vertical, dashed line
denotes the Fermi energy. Note that the total DOS in the figure corresponds
to per two atoms. \label{fig:dos-fese-rl&fete-h}}

\end{figure}

\begin{figure}
\includegraphics[clip,scale=0.5]{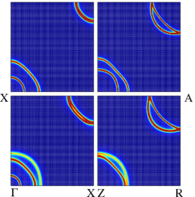}

\caption{(Color online) The Fermi surface of FeSe (top) and FeTe (bottom) in
$\Gamma$-X-M and Z-R-A planes obtained using experimental lattice
parameters and theoretically relaxed $z_{Se}$. \label{fig:fs-fese-rl-fete-h}}

\end{figure}

To be able to separate out the effects of disorder from the atomic-sphere-approximation-related
changes, in this section we compare the electronic structure of ordered
FeSe and FeTe alloys with the more accurate full-potential results.
In Figs. \ref{fig:bs-fese-fete}-\ref{fig:fs-fese-rl-fete-h}, we
show the band structure, the density of states and the Fermi surface
of FeSe and FeTe alloys, which are similar to that of Ref. \cite{ASubedi}
except for the gap between Se- and Fe-derived bands around 2.5 eV
below $E_{F}$ in FeSe. Since it is already known that electronic
structure calculations of FeSe are very sensitive to the height of
the Se-layer above the Fe-plane, we have carried out full-potential,
linear muffin-tin orbital (FP-LMTO) calculations for FeSe using the
lattice parameters as above. We find that the gap predicted by FP-LMTO
is $\approx$0.3 eV, which increases to $\approx$0.4 eV if the muffin-tin
spheres are enlarged to atomic-spheres within the full-potential approach.
In the atomic-sphere approximation, which uses additional approximations,
including that of making the potential spherical, the gap is found
to be $\approx$0.8 eV. However, the Fe-derived bands, which are responsible
for superconductivity in these alloys, compare well with the more
accurate FP-LMTO calculations as well as with that of Ref. \cite{ASubedi}.

Using the idea of FS nesting, it is possible to get some quantitative
measure of the response of the system without evaluating the susceptibility.
For example, if the FS around $\Gamma$ point matches exactly with
the FS around M point when displaced by a reciprocal space vector
then the nesting is optimal. In the present context, the optimal nesting
corresponds to FS at $\Gamma$ and M points having matching radii
and sharp Fermi surfaces (reflected by thin red lines in the figures).
Any deviation, either from the matching radii or sharpness of the
Fermi surfaces (reflected by diffused and/or broadened lines in the
figures), generally, reduces the effect of nesting. 

The FS of FeSe, shown in Fig. \ref{fig:fs-fese-rl-fete-h}, consists
of two hole-like sheets around $\Gamma$ point and two electron-like
sheets around M point, both sheets being derived from $xz$ ($yz$)
and $xy$ bands of Fe. In FeTe, there are three hole-like sheets around
$\Gamma$ point and only two sheets around M point as shown in Fig.
\ref{fig:fs-fese-rl-fete-h}. The Fermi surfaces shown in Fig. \ref{fig:bs-fese-fete}
reveal enhanced FS nesting at $\Gamma$-X-M plane in FeSe than in
FeTe. Note that the two bands around M point in FeSe are not resolved
in Fig. \ref{fig:bs-fese-fete}. In the following, we will see if
the changes in the shape and the nesting of the FS of FeSe induced
upon alloying can be used to understand the changes in the superconducting
properties of these alloys within the framework of the spin-fluctuation
theories.

\begin{figure}
\includegraphics[clip,scale=0.5]{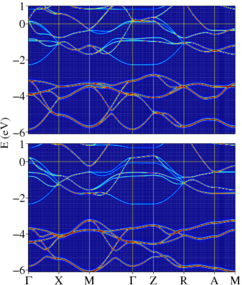}

\caption{(Color online) The band structure of tetragonal FeSe along the BZ
symmetry directions using experimental lattice parameters with $z_{Se}$
equal to its experimental (top) or theoretically relaxed (bottom)
value. In both the calculations, the exchange-correlation potential
was parametrized within the generalized-gradient approximation as
described in the text. The Fermi energy is indicated by the horizontal
line at $0$ eV. \label{fig:bs-fese&fese-r}}

\end{figure}

\begin{figure}
\includegraphics[clip,scale=0.3]{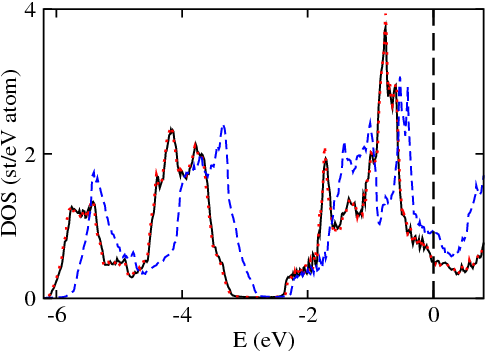}\caption{(Color online) The total DOS of tetragonal FeSe with the experimental
lattice parameters, including experimental $z_{Se}$ (blue, dash),
and with theoretical $z_{Se}$ (red, dot) calculated within the generalized-gradient
approximation. The total DOS of FeSe (black, solid) as obtained in
Fig. \ref{fig:dos-fese-rl&fete-h} is also shown. \label{fig:dos-fese-&fese-r&fese-rl}}

\end{figure}

\section{Effects of theoretical \emph{$vs$}. experimental $z$$_{Se}$\label{sec:Effects-of-zse}}

The role of Se height above the Fe plane, given by $z_{Se}$, plays
a crucial role in determining the electronic properties, especially
the magnetic properties \cite{ZPYin}, of FeSe and its alloys. It
is not surprising that $z_{Se}$ plays such an important role in deciding
the electronic properties of FeSe because a change in $z_{Se}$ directly
impacts the Fe $d$-orbitals which, in turn, affect the Fe-derived
$d$-bands around $E_{F}$. Most of the theoretical work on FeSe have
used the experimental values of the lattice parameters $a$ and $c$,
and theoretically relaxed value for $z_{Se}$. To see what role does
$z_{Se}$ play in the present context, we have studied FeSe and its
alloys using the experimental value of $z_{Se}=0.266$ and the theoretically
relaxed value of $z_{Se}=0.2343$ \cite{ASubedi}.

In Fig. \ref{fig:bs-fese&fese-r}, we show the band structure of FeSe
calculated with the experimental as well as the theoretically relaxed
value of $z_{Se}$. In both cases, the experimental values of the
lattice parameters $a$ and $c$ were used, and the exchange-correlation
potential was parametrized using the generalized-gradient approximation
of Perdew \emph{et al}. \cite{perdew2}. The use of experimental $z_{Se}$,
which is larger than the theoretical value, allows most of the bands,
including the Se-derived bands, to move up with respect to $E_{F}$.
Such a movement results in three bands crossing $E_{F}$ along  $\Gamma$-X,
M-$\Gamma$ and Z-R. The use of theoretical $z_{Se}$, which reduces
the Se-height above the Fe plane, affects the individual Fe-derived
$d$-bands around $E_{F}$ differently due to the orientation of the
$d$-orbitals. In particular, we find that the $3z^{2}-1$ derived
band along $\Gamma$-X and $\Gamma$-Z is suppressed more than the
other two bands, resulting in only two bands crossing $E_{F}$ along
 $\Gamma$-X, M-$\Gamma$ and Z-R for theoretically relaxed $z_{Se}$. 

The effects of using experimental or theoretical $z_{Se}$ get amplified
in the total DOS as shown in Fig. \ref{fig:dos-fese-&fese-r&fese-rl}.
The higher value of $z_{Se}$ shifts the DOS towards $E_{F}$ as well
as redistributes some of the electronic states between 0 and -2 eV.
The Fe-peak in the DOS is closer to $E_{F}$ for experimental $z_{Se}$
than for the theoretical $z_{Se}$. From Fig. \ref{fig:dos-fese-&fese-r&fese-rl},
we also find that the DOS of FeSe obtained with the Barth-Hedin exchange-correlation
potential is essentially identical to the DOS obtained with Perdew-Wang
\cite{perdew1} exchange-correlation potential within the local-density
approximation. 

We have also calculated the Fermi surface of FeSe with experimental
and theoretical $z_{Se}$. As discussed above, in the context of the
band structure of FeSe, we find three bands around $\Gamma$ and Z
points for experimental $z_{Se}$.

\section{Electronic structure of disordered $Fe_{1+x}Se$ and $FeSe_{1-x}$
alloys}

Generally, the synthesis of FeSe results in an alloy with either excess
of Fe or deficiency in Se \cite{TMMcQueen,EPom,AJWilliams}. Measurements
on such alloys have shown that Fe$_{1+\delta}$Se is superconducting
 only when $\delta\leq0.01$ and the underlying lattice is orthorhombic
\cite{TMM2}. On the other hand, Se-deficient alloys FeSe$_{1-x}$
remain superconducting for $x\leq0.18$ in the tetragonal structure
\cite{FCHsu,TMMcQueen}. In order to understand the contrasting superconducting
properties of Fe$_{1+\delta}$Se and FeSe$_{1-x}$ alloys, we have
studied their normal state electronic structure as a function of $\delta$
and $x$ with $0\leq\delta\leq0.06$ and $0\leq x\leq0.18$.

\begin{figure}
\subfloat[]{\includegraphics[clip,scale=0.5]{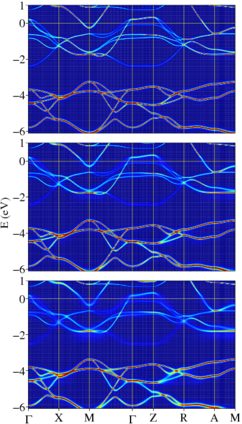}

}\subfloat[]{\includegraphics[clip,scale=0.5]{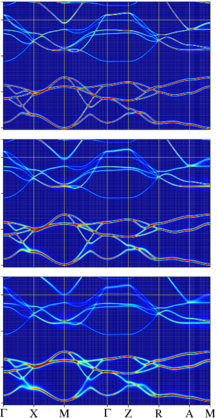}}

\caption{(Color online) The band structure of (a) Fe$_{1+\delta}$Se and (b)
FeSe$_{1-x}$ with $\delta$ and $x$$=0.01$ (top), $0.03$ (middle)
and $0.06$ (bottom) along the BZ symmetry directions. The Fermi energy
is indicated by the horizontal line at $0$ eV.\label{fig:bs-fe01-03-06se&fese01-03-06}}

\end{figure}

\begin{figure}
\subfloat[]{\includegraphics[scale=0.25]{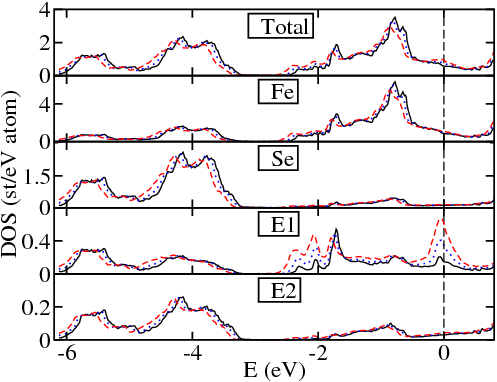}

}\subfloat[]{\includegraphics[clip,scale=0.25]{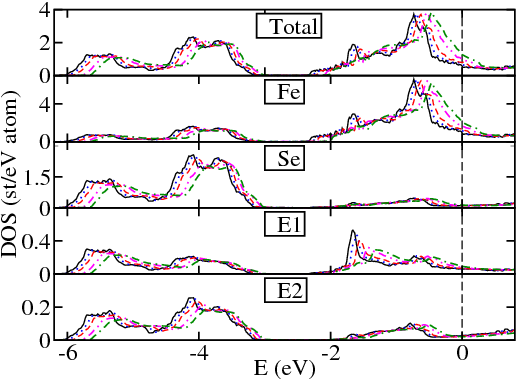}}

\subfloat[]{\includegraphics[clip,scale=0.28]{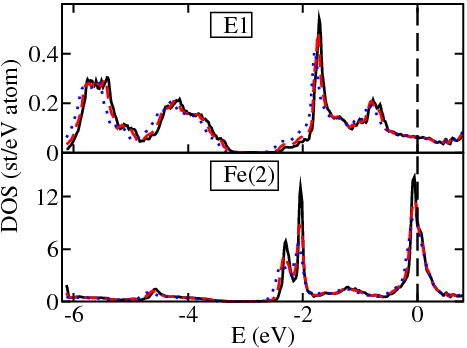}

}\subfloat[]{\includegraphics[clip,scale=0.28]{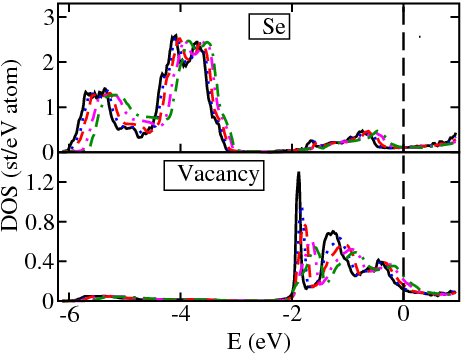}

}

\caption{(Color online) The total and the sub-lattice-resolved DOS of (a) Fe$_{1+\delta}$Se
and (b) FeSe$_{1-x}$ with $\delta$ and $x$$=0.01$ (black, solid),
$0.03$ (blue, dot) and $0.06$ (red, dash). For FeSe$_{1-x}$, the
DOS corresponding to $x$$=0.12$ (magenta, double-dot dash) and $0.18$
(green, double-dash dot) are also shown. The vertical, dashed line
denotes the Fermi energy. The atom-resolved DOS at (c) the E1-site
of Fe$_{1+\delta}$Se and (d) the Se-site of FeSe$_{1-x}$ are also
shown. \label{fig:dos-fe01-03-06se&fese01-03-06}}

\end{figure}

\begin{figure}
\includegraphics[clip,scale=0.4]{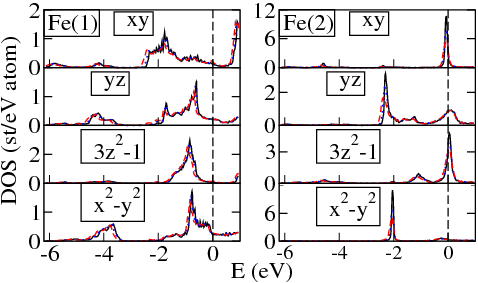}\caption{(Color online) The $d$-resolved DOS at the Fe(1) (left panel) and
Fe(2) (right panel) sites in Fe$_{1+\delta}$Se with $\delta=0.01$
(black, solid), $0.03$ (blue, dot), and $0.06$ (red, dash). The
vertical, dashed line denotes the Fermi energy. \label{fig:dos-fe01-03se-fe1-fe2-d}}

\end{figure}

\begin{figure}
\subfloat[]{\includegraphics[clip,scale=0.5]{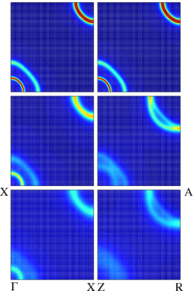}

}\subfloat[]{\includegraphics[clip,scale=0.5]{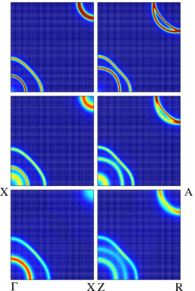}

}\caption{(Color online) The Fermi surface of (a) Fe$_{1+\delta}$Se and (b)
FeSe$_{1-x}$ with $\delta$ and $x$$=0.01$ (top), $0.03$ (middle)
and $0.06$ (bottom) in $\Gamma$-X-M and Z-R-A planes. \label{fig:fs-fe01-03-06se&fese01-03-06}}

\end{figure}

The changes in the band structure, the density of states and the FS
in Fe$_{1+\delta}$Se and FeSe$_{1-x}$ alloys with increasing $\delta$
and $x$ are shown in Figs. \ref{fig:bs-fe01-03-06se&fese01-03-06}-\ref{fig:fs-fe01-03-06se&fese01-03-06}.
In the unit cell, the excess Fe atom, denoted by Fe(2) (The Fe at
the stoichiometric site is denoted by Fe(1)), is kept in the Se plane
with $z_{Fe}=-z_{Se}$, which is one of the two in-equivalent sub-lattices
containing  empty spheres. Our main conclusions are independent of
the possible locations of the excess Fe atoms within the FeSe lattice,
as was found in Ref. \cite{TMMcQueen}. The increase in disorder induced
by the presence of excess Fe can be clearly seen, as indicated by
the diffused intensity, in the Fe-derived bands in Fig. \ref{fig:bs-fe01-03-06se&fese01-03-06}(a).
The Se-derived bands are left relatively untouched by the excess Fe.
From the DOS of Fe$_{1+\delta}$Se alloys, shown in Figs. \ref{fig:dos-fe01-03-06se&fese01-03-06}(a)
and (c), and Fig. \ref{fig:dos-fe01-03se-fe1-fe2-d}, we find that
the excess Fe kept on the E1 sub-lattice creates states around $E_{F}$
and around $-2$ eV. The possibility of the excess Fe becoming magnetic
due to the increased local density of states at $E_{F}$ can be clearly
seen in the $d$-resolved DOS of Fe(1) and Fe(2) in Fig. \ref{fig:dos-fe01-03se-fe1-fe2-d}.
On the other hand, the increase in excess Fe quickly destroys the
FS nesting as seen from Fig. \ref{fig:fs-fe01-03-06se&fese01-03-06}(a).
Note that even for $\delta=0.01,$ the nesting is not optimal. 

For Fe$_{1.01}$Se in the orthorhombic structure, the overall change
in electronic structure with respect to the tetragonal structure is
symmetry-induced and minimal. There are three bands crossing $E_{F}$
around $\Gamma$ point instead of two bands as is the case in the
tetragonal structure, leading to some changes in the FS around $\Gamma$
point. The DOS remains essentially unchanged with respect to the tetragonal
case.

The FeSe$_{1-x}$ alloys remain superconducting over a wide range
of $x-$values, indicating an unusually small impact on the bands
around $E_{F}$ due to vacancies in the alloy. Our results for FeSe$_{1-x}$
alloys, shown in Figs. \ref{fig:bs-fe01-03-06se&fese01-03-06}-\ref{fig:fs-fe01-03-06se&fese01-03-06}
seem to confirm it. The vacancies on the Se sub-lattice affect the
Se-derived bands which are away from $E_{F}$, leaving the bands around
$E_{F}$ essentially untouched for small $x$. In addition to the
lowering of $E_{F}$ due to electron loss, for $x\geq0.06$ the bands
around $E_{F}$ begin to get affected by disorder. For example, for
$x\geq0.12$, we find that the Se-derived bands have been substantially
diffused. The loss of electron due to increasing Se vacancy moves
$E_{F}$ inward, seen clearly in Fig. \ref{fig:dos-fe01-03-06se&fese01-03-06}(b),
bringing the peak in DOS due to the stoichiometric Fe closer to $E_{F}$
as shown in Fig. \ref{fig:dos-fe01-03-06se&fese01-03-06}(d), and
thereby increasing the total DOS at $E_{F}$. The largest contribution
to the DOS at $E_{F}$ comes from the states of Fe(1) having $x^{2}-y^{2}$
and $xy$ symmetries. Such an increase in the DOS at $E_{F}$ may
lead to magnetism in the alloy \cite{KWLee}. We emphasize that Fe(2)
atoms are responsible for possible magnetism in Fe$_{1+\delta}$Se,
while Fe(1) atoms may lead to magnetic behavior in FeSe$_{1-x}$.
With increasing $x$, the gradual loss of possible FS nesting in Fe$_{1+\delta}$Se
and FeSe$_{1-x}$ can also be seen in Fig. \ref{fig:fs-fe01-03-06se&fese01-03-06}.
If FS nesting is crucial for superconductivity in FeSe$_{1-x}$ alloys
then clearly the possibility of superconductivity is diminished for
large $x$ as the nesting is essentially destroyed as shown in Fig.
\ref{fig:fs-fe01-03-06se&fese01-03-06}. However, experiments \cite{EPom,AJWilliams}
have shown that FeSe$_{1-x}$ for large $x$, contain impurity phases
such as elemental Fe, leading to incorrect determination of Se content.
Therefore, the Se content in superconducting FeSe$_{1-x}$ alloys
must be determined carefully. 

In both Fe$_{1+\delta}$Se and FeSe$_{1-x}$, our results clearly
show that the effects of excess Fe or Se deficiency on electronic
structure of FeSe are very different, and that for $x\geq0.06$ the
FS nesting is essentially destroyed.

\section{Energetics of excess $Fe$ and $Se$ deficiency in $FeSe$}

To get an estimate of the energetics involved in the formation of
FeSe alloys with either excess Fe or Se deficiency, we have calculated
the formation energy of Fe$_{1+\delta}$Se and FeSe$_{1-x}$ alloys
as described in Sec. \ref{sec:Computational-Method}. In Fe$_{1+\delta}$Se,
we find that the formation energy $E_{form}^{e}=$ -98, -94 and -87
meV/atom for $\delta=$ 0.01, 0.03 and 0.06, respectively. The vacancy
formation energy in FeSe$_{1-x}$ alloys is found to be $E_{form}^{v}=$115,
126, 132, 134 and 136 meV/atom for $x=$ 0.01, 0.03, 0.06, 0.12, and
0.18, respectively. As our calculations are done within the atomic-sphere
approximation and without taking relaxation into account, these results
are expected to change with the incorporation of full-potential and
relaxation effects \cite{KWLee}.

\begin{figure}
\includegraphics[bb=54bp 24bp 703bp 533bp,clip,scale=0.35]{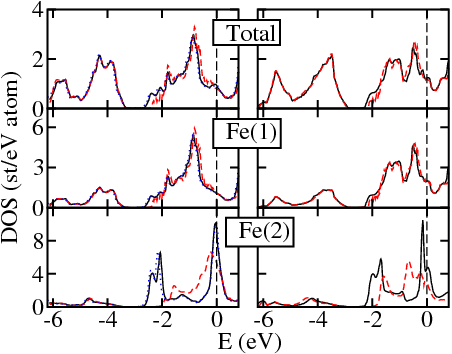}

\caption{(Color online) The total and the Fe(1) and Fe(2) DOS in Fe$_{1.06}$Se
with theoretically relaxed $z_{Se}$ (left panel) and experimental
$z_{se}$ (right panel). The DOS correspond to different positions
of the excess Fe (Fe(2)) given by $z_{Fe}=-z_{Se}$ (black, solid)
with Fe(2) in the Se-plane, $z_{Fe}=0.7755$ (blue, dot) with Fe(2)
slightly below the Se-plane and $z_{Fe}=0.5$ (red, dash) with Fe(2)
in the middle of the two Se-planes. The vertical, dashed line denotes
the Fermi energy. \label{fig:dos-fe06-m-x}}

\end{figure}

In order to check the effects of keeping the excess Fe in Fe$_{1+\delta}$Se
at the various interstitial regions, we have studied Fe$_{1.06}$Se
alloy keeping the excess Fe (Fe(2)) in the Se-plane with $z_{Fe}=-z_{Se}$,
slightly below the Se-plane with $z_{Fe}=0.7755$ and in the middle
of the two Se-planes with $z_{Fe}=0.5$ for theoretically-relaxed
$z_{Se}$ as well as experimental $z_{se}$. As noted in Sec. \ref{sec:Effects-of-zse},
the DOS of FeSe and its alloys are expected to be different for theoretically-relaxed
$z_{Se}$ and experimental $z_{se}$, as is the case for Fe$_{1.06}$Se
in Fig. \ref{fig:dos-fe06-m-x}. Not surprisingly, the DOS corresponding
to $z_{Fe}=-z_{Se}$ and $z_{Fe}=0.7755$ are essentially identical
for theoretically relaxed $z_{Se}$. However, when the excess Fe is
placed in between the two Se-planes instead of in the Se-plane, the
resulting Fe(2) DOS's are somewhat different from each other, as shown
in the bottom panel of Fig. \ref{fig:dos-fe06-m-x}. These differences
arise due to the relatively free space available to Fe(2) when $z_{Fe}=0.5$.
We also find that it is only with the inclusion of the Madelung potential
in the Muffin-Tin-corrected total energy that Fe(2) prefers to go
to $z_{Fe}=-z_{Se}$ instead of $z_{Fe}=0.5$ by 11 meV/atom for experimental
$z_{Se}$. We must point out that a more reliable way of calculating
the site-preference involving different structures is through the
full-potential approach with relaxation included.

\section{Electronic structure of disordered $FeSe_{0.9}S_{0.1}$, $FeSe_{1-y}Te_{y}$,
$Fe_{0.9}Co_{0.1}Se$, $Fe_{0.9}Ni_{0.1}Se$ and $Fe_{0.91}Cu_{0.1}Se$
alloys}

\begin{figure}
\subfloat[]{\includegraphics[clip,scale=0.5]{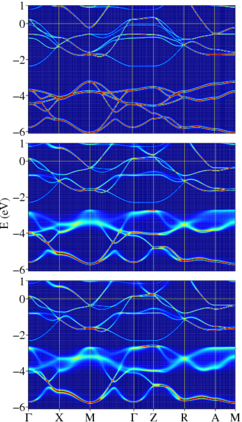}}\subfloat[]{\includegraphics[clip,scale=0.5]{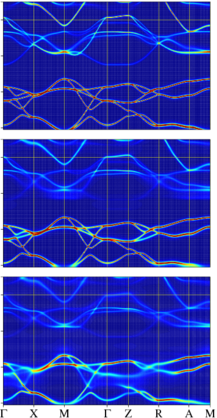}}\caption{(Color online) The band structure of (a) FeSe$_{0.9}$S$_{0.1}$ (top),
FeSe$_{0.75}$Te$_{0.25}$ (middle), FeSe$_{0.5}$Te$_{0.5}$(bottom),
and (b) Fe$_{0.9}$Co$_{0.1}$Se (top), Fe$_{0.9}$Ni$_{0.1}$Se (middle)
and Fe$_{0.91}$Cu$_{0.1}$Se (bottom) along the BZ symmetry directions.
The Fermi energy is indicated by the horizontal line at $0$ eV.\label{fig:bs-fese-s-te-te&fe-co-ni-cu-se}}

\end{figure}

\begin{figure}
\subfloat[]{\includegraphics[clip,scale=0.25]{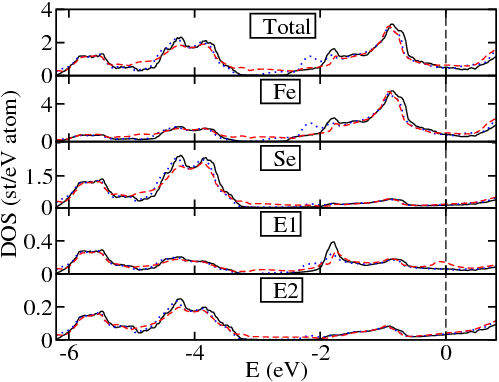}

}\subfloat[]{\includegraphics[clip,scale=0.25]{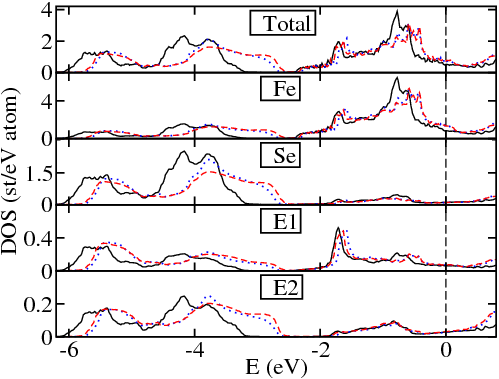}}

\subfloat[]{\includegraphics[clip,scale=0.28]{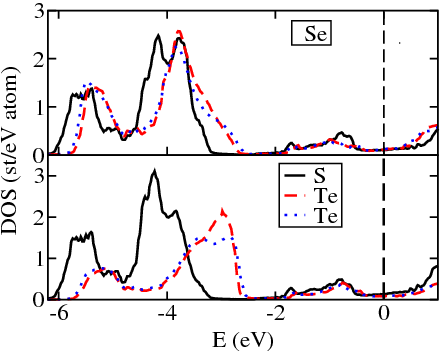}}\subfloat[]{\includegraphics[clip,scale=0.28]{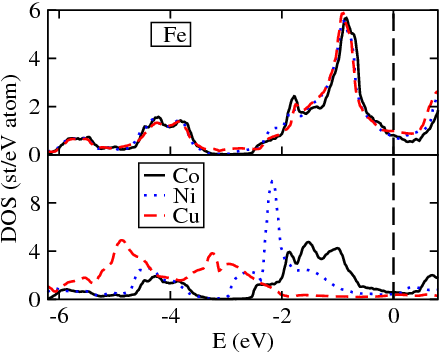}}\caption{(Color online) The total and the sub-lattice-resolved DOS of (a) S
and Te substitutions on Se sub-lattice and (b) Co, Ni and Cu substitutions
on Fe sub-lattice. The DOS in (a) correspond to FeSe$_{0.9}$S$_{0.1}$
(black, solid), FeSe$_{0.75}$Te$_{0.25}$ (blue, dot), FeSe$_{0.5}$Te$_{0.5}$
(red, dash), and in (b) Fe$_{0.9}$Co$_{0.1}$Se (black, solid), Fe$_{0.9}$Ni$_{0.1}$Se
(blue, dot) and Fe$_{0.91}$Cu$_{0.1}$Se (red, dash). The vertical,
dashed line denotes the Fermi energy. The atom-resolved, DOS of S
and Te substitutions at the Se sub-lattice and Co, Ni and Cu substitutions
at the Fe sub-lattice are shown in (c) and (d), respectively. \label{fig:dosfe-s-se-te&fe-co-ni-cu-se}}

\end{figure}

\begin{figure}
\subfloat[]{\includegraphics[scale=0.5]{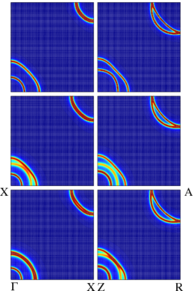}

}\subfloat[]{\includegraphics[scale=0.5]{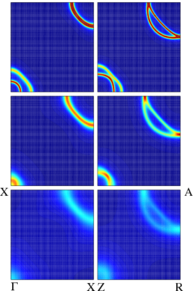}

}\caption{(Color online) The Fermi surface of (a) FeSe$_{0.9}$S$_{0.1}$ (top),
FeSe$_{0.75}$Te$_{0.25}$ (middle), FeSe$_{0.5}$Te$_{0.5}$ (bottom),
and (b) Fe$_{0.9}$Co$_{0.1}$Se (top), Fe$_{0.9}$Ni$_{0.1}$Se (middle)
and Fe$_{0.91}$Cu$_{0.1}$Se (bottom) in $\Gamma$-X-M and Z-R-A
planes.\label{fig:fs-fese-s-te-te&fe-co-ni-cu-se}}

\end{figure}

To understand the changes in the superconducting properties of FeSe
upon substitution of impurities through the changes in the normal
state electronic properties, we have studied the effects of substituting
S and Te on Se sub-lattice and Co, Ni and Cu on Fe sub-lattice. In
Figs. \ref{fig:bs-fese-s-te-te&fe-co-ni-cu-se}-\ref{fig:fs-fese-s-te-te&fe-co-ni-cu-se},
we show our results for the band structure, density of states and
FS of FeSe$_{0.9}$S$_{0.1}$, FeSe$_{0.75}$Te$_{0.25}$, FeSe$_{0.5}$Te$_{0.5}$,
Fe$_{0.9}$Co$_{0.1}$Se, Fe$_{0.9}$Ni$_{0.1}$Se and Fe$_{0.91}$Cu$_{0.1}$Se
alloys. Based on the atomic size-mismatch, it is expected that the
substitution of Te would lead to more disorder than the substitution
of S on Se sub-lattice in FeSe. Indeed, we find that the effects of
disorder are minimal due to S but substitution of Te creates states
in the gap region with substantial disorder in the Se-derived bands.
The states created in the gap between Se- and Fe-derived bands can
be clearly seen in Fig. \ref{fig:bs-fese-s-te-te&fe-co-ni-cu-se}(a)
and (c). Surprisingly, for FeSe$_{0.5}$Te$_{0.5}$ alloy, we find
that the substitution of Te has rearranged the bands around $E_{F}$
such that the possibility of FS nesting is maximized as shown in the
bottom panel of Fig. \ref{fig:fs-fese-s-te-te&fe-co-ni-cu-se}(a).

The presence of Co, Ni and Cu in Fe sub-lattice of FeSe is expected
to affect the bands around $E_{F}$ as well as move $E_{F}$ up due
to disorder and electron doping, thereby significantly changing its
superconducting properties. As shown in Fig. \ref{fig:bs-fese-s-te-te&fe-co-ni-cu-se}(b),
a $10\%$ addition of Ni or Cu moves $E_{F}$ above the two bands
around $\Gamma$ point. The substitution of Cu seems to disorder both
the Fe- and the Se-derived bands, as can be seen from the diffused
intensity of the bands in Fig. \ref{fig:bs-fese-s-te-te&fe-co-ni-cu-se}(b).
The addition of Co, Ni or Cu creates states in the gap region around
-2 eV as seen from Fig. \ref{fig:dosfe-s-se-te&fe-co-ni-cu-se}(b)
and (d). The combined effect of disorder and electron addition diminishes
the possibility of FS nesting in Fe$_{0.9}$Co$_{0.1}$Se, and destroys
it in Fe$_{0.9}$Ni$_{0.1}$Se and Fe$_{0.91}$Cu$_{0.1}$Se as seen
in Fig. \ref{fig:fs-fese-s-te-te&fe-co-ni-cu-se}(b).

\section{Describing disorder : Rigid-band, Virtual-crystal, Supercell and
Coherent-potential approximations}

In order to emphasize the importance of describing the effects of
substitutional disorder accurately and reliably in metals, in general,
and in FeSe and its alloys, in particular, we have studied selected
FeSe alloys using the rigid-band approximation, the virtual-crystal
approximation and the supercell approximation. In this section, we
describe and compare our results, in terms of DOS and Fermi surface,
of the rigid-band approximation, the virtual-crystal approximation
and the supercell approximation with that of the coherent-potential
approximation for the selected FeSe alloys.

\begin{figure}
\includegraphics[scale=0.4]{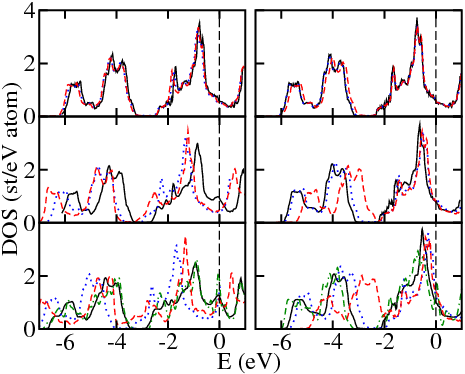}\caption{(Color online) The total DOS of Fe$_{1+\delta}$Se (left panel) and
FeSe$_{1-x}$ (right panel) with $\delta$ and $x$$=0.01$ (top),
$0.06$ (middle) and $0.125$ (bottom) calculated in the rigid-band
approximation (blue, dot), the virtual-crystal approximation (red,
dash) and the coherent-potential approximation (black, solid). In
the bottom panel, the total DOS obtained in the supercell approximation
(green, double-dash dot) using Fe$_{9}$Se$_{8}$ (left) and Fe$_{8}$Se$_{7}$
(right) are also shown. The vertical, dashed line denotes the Fermi
energy. \label{fig:dos-fe1-12se&fese1-12-rl-b-v-sc}}

\end{figure}

\begin{figure}
\includegraphics[scale=0.25]{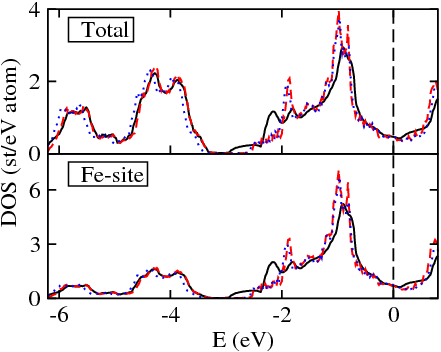}

\caption{(Color online) The total and the virtual-atom-site DOS of Fe$_{0.9}$Ni$_{0.1}$Se
(red, dash), calculated in the virtual-crystal approximation as described
in the text. The corresponding DOS in the rigid-band approximation
(blue, dot), and the coherent-potential approximation (black, solid)
are also shown. The vertical, dashed line denotes the Fermi energy.
\label{fig:dos-fesete&fenise-v}}

\end{figure}

We have studied Fe$_{1+\delta}$Se and FeSe$_{1-x}$ with $\delta$
and $x$$=0.01$, $0.06$ and $0.125$, and Fe$_{0.9}$Ni$_{0.1}$Se
alloys in both the rigid-band approximation and the virtual-crystal
approximation. For $\delta$ and $x=0.125$, studied for comparison
with the supercell approximation, we have used the lattice parameters
corresponding to $\delta=0.06$ and $x=0.12$, respectively. We have
also studied Fe$_{9}$Se$_{8}$ and Fe$_{8}$Se$_{7}$ as well as
Fe$_{8}$Se$_{8}$ in the supercell approximation. The selected alloys
cover a wide range of systems with effects due to disorder involving
addition of $d$-electrons in Fe$_{1+\delta}$Se, removal of $s$-
and $p$-electrons in FeSe$_{1-x}$, and similar atoms in Fe$_{0.9}$Ni$_{0.1}$Se.
Generally, one expects the rigid-band approximation to work well for
free-electron-like systems or systems consisting of similar atoms.
On the other hand, virtual-crystal approximation is expected to work
well for systems consisting of atoms from nearby columns of the periodic
table such as Fe$_{0.9}$Ni$_{0.1}$Se. The supercell approximation
or its more efficient version known as the special quasi-random structure
\cite{AZunger} for describing the disordered alloy can be used to
extract many properties of alloys such as the density of states and
phase stability. 

In Fig. \ref{fig:dos-fe1-12se&fese1-12-rl-b-v-sc}, we compare the
total DOS of Fe$_{1+\delta}$Se and FeSe$_{1-x}$ with $\delta$ and
$x$$=0.01$, $0.06$ and $0.125$ calculated in the rigid-band and
virtual-crystal approximations with that of the coherent-potential
approximation. For $\delta$ and $x$$=0.01$, we find that the total
DOS obtained in the rigid-band approximation is close to that of the
coherent-potential approximation. For higher concentrations, the rigid-band
DOS begins to differ substantially from the CPA DOS for Fe$_{1+\delta}$Se.
Not surprisingly, in FeSe$_{1-x}$ the Se vacancy, involving the absence
of $s$ and $p$ electrons, leads to a relatively small disagreement
up to $x=0.06$.

For both Fe$_{1+\delta}$Se and FeSe$_{1-x}$, the DOS calculated
with the virtual-crystal approximation differs substantially with
the CPA DOS for $\delta$ and $x$$\geq0.06$, as can be seen from
Fig. \ref{fig:dos-fe1-12se&fese1-12-rl-b-v-sc}. Note that for $x$$=0.125$,
the virtual-crystal approximation replaces the Se atom on the Se sub-lattice
in FeSe$_{1-x}$ by a virtual atom with atomic number equal to 33.25.
Thus, we find that the virtual-crystal approximation essentially fails
to describe accurately the effects of disorder in the DOS of either
Fe$_{1+\delta}$Se or FeSe$_{1-x}$ for $\delta$ and $x\geq0.06$.
However, the virtual-crystal approximation is expected to work better
if the virtual atom is made out of atoms with similar atomic numbers
as in Fe$_{0.9}$Ni$_{0.1}$Se. In Fig. \ref{fig:dos-fesete&fenise-v},
we compare the total and the Fe sub-lattice DOS in Fe$_{0.9}$Ni$_{0.1}$Se
in the virtual-crystal approximation with the corresponding CPA DOS.
Indeed, we find a good agreement in the DOS of Fe$_{0.9}$Ni$_{0.1}$Se.
Some of the differences in the DOS of Fe$_{0.9}$Ni$_{0.1}$Se around
-1.9 eV and -1.0 eV highlight one of the shortcomings of the virtual-crystal
approximation (as well as that of the supercell approximation) in
assuming a periodic lattice with no energy-dependent electron scattering.
The virtual-crystal approximation, as implemented here, cannot be
applied to FeSe$_{0.5}$Te$_{0.5}$. A more appropriate approach for
implementing the virtual-crystal approximation in alloys may be through
the use of psuedopotentials \cite{DVanderbilt}.

In Fig. \ref{fig:dos-fe1-12se&fese1-12-rl-b-v-sc}, the total DOS
obtained for Fe$_{9}$Se$_{8}$ and Fe$_{8}$Se$_{7}$, corresponding
to the supercell approximation for 0.125 of excess Fe and 0.125 of
Se vacancy per atom in FeSe, respectively, are also shown. The total
DOS for Fe$_{9}$Se$_{8}$ is in overall agreement with the corresponding
CPA DOS of Fe$_{1.125}$Se. For Se vacancy, the total DOS of Fe$_{8}$Se$_{7}$
is very close to the fully-relaxed, full-potential supercell result
of Ref. \cite{KWLee}, as shown in their Fig. 2. A comparison of the
supercell DOS of Fe$_{8}$Se$_{7}$ with the CPA DOS of FeSe$_{0.875}$
shows some differences around Se-region and around $E_{F}$. Specially,
the sharp peak close to $E_{F}$ is missing in the CPA DOS. If the
presence of the peak is confirmed by experiment, then it would indicate
the need to go beyond the single-site CPA. The supercell approximation
does provide the possibility of including several atomic environments,
however, its inherent flaw of not including energy-dependent electron
scattering limits its applicability. 

\begin{figure}
\subfloat[]{\includegraphics[scale=0.5]{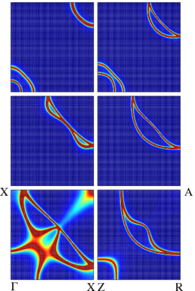}

}\subfloat[]{\includegraphics[scale=0.5]{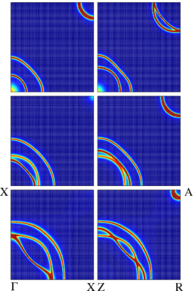}}\caption{(Color online) The Fermi surface, calculated in the rigid-band approximation
as described in the text, of (a) Fe$_{1+\delta}$Se and (b) FeSe$_{1-x}$
with $\delta$ and $x$$=0.01$ (top), $0.06$ (middle) and $0.125$
(bottom) in $\Gamma$-X-M and Z-R-A planes. \label{fig:fs-fe01-06-12se&fese01-06-12-rb}}

\end{figure}

\begin{figure}
\subfloat[]{\includegraphics[scale=0.5]{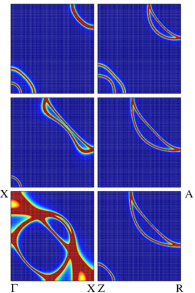}

}\subfloat[]{\includegraphics[scale=0.5]{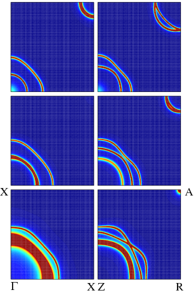}}\caption{(Color online) The Fermi surface, calculated in the virtual-crystal
approximation as described in the text, of (a) Fe$_{1+\delta}$Se
and (b) FeSe$_{1-x}$ with $\delta$ and $x$$=0.01$ (top), $0.06$
(middle) and $0.125$ (bottom) in $\Gamma$-X-M and Z-R-A planes.
\label{fig:fs-fe01-06-12se&fese01-06-12-v}}

\end{figure}

\begin{figure}
\subfloat[]{\includegraphics[scale=0.5]{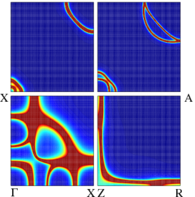}

}\subfloat[]{\includegraphics[scale=0.5]{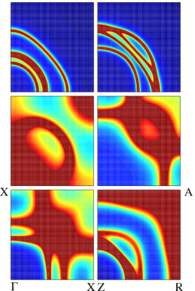}}\caption{(Color online) The Fermi surface of (a) Fe$_{0.9}$Ni$_{0.1}$Se,
calculated in the rigid-band approximation (top) and the virtual-crystal
approximation (bottom) in $\Gamma$-X-M and Z-R-A planes. In (b) the
Fermi surfaces of Fe$_{8}$Se$_{8}$ (top), Fe$_{9}$Se$_{8}$ (middle)
and Fe$_{8}$Se$_{7}$ (bottom) in $\Gamma$-X-M and Z-R-A planes,
calculated using the supercell approximation, are shown. \label{fig:fs-fe90ni10se-rb-v-fe8-9se7}}

\end{figure}

The $\mathbf{k}$-resolved properties of alloys provide a more stringent
test for the accuracy and the reliability of the various approximations
for describing disorder that we have discussed so far. Therefore,
we have calculated the Fermi surface of the selected alloys in the
$\Gamma$-X-M and Z-R-A planes in the rigid-band, virtual-crystal
and supercell approximations, and the results are shown in Figs. \ref{fig:fs-fe01-06-12se&fese01-06-12-rb}-\ref{fig:fs-fe90ni10se-rb-v-fe8-9se7}.
The Fermi surface of Fe$_{1+\delta}$Se and FeSe$_{1-x}$ obtained
in the rigid-band approximation, Fig. \ref{fig:fs-fe01-06-12se&fese01-06-12-rb},
agrees with the corresponding CPA Fermi surface, Fig. \ref{fig:fs-fe01-03-06se&fese01-03-06},
for $\delta$ and $x$$=0.01$ only. The Fermi surface obtained in
the virtual-crystal approximation, shown in Fig. \ref{fig:fs-fe01-06-12se&fese01-06-12-v},
disagrees with both the rigid-band, Fig. \ref{fig:fs-fe01-06-12se&fese01-06-12-rb},
and the CPA, Fig. \ref{fig:fs-fe01-03-06se&fese01-03-06}, results
for $\delta$ and $x\geq0.06$. Similarly, for Fe$_{0.9}$Ni$_{0.1}$Se
the Fermi surface in the rigid-band approximation is closer to the
corresponding CPA Fermi surface than the Fermi surface obtained in
the virtual-crystal approximation as shown in Fig. \ref{fig:fs-fe90ni10se-rb-v-fe8-9se7}(a).
Finally, we show the Fermi surface of Fe$_{8}$Se$_{8}$, Fe$_{9}$Se$_{8}$
and Fe$_{8}$Se$_{7}$ in Fig. \ref{fig:fs-fe90ni10se-rb-v-fe8-9se7}(b),
obtained in the supercell approximation. Due to the folding in $k_{x}$
and $k_{y}$ directions, the Fermi surface of Fe$_{9}$Se$_{8}$ and
Fe$_{8}$Se$_{7}$ cannot be compared directly with the CPA Fermi
surface obtained using the primitive cell. Therefore, we also show
in Fig. \ref{fig:fs-fe90ni10se-rb-v-fe8-9se7}(b), the Fermi surface
of Fe$_{8}$Se$_{8}$. We find that the supercell approximation of
Fe$_{9}$Se$_{8}$ and Fe$_{8}$Se$_{7}$ predict a very different
Fermi surface of disordered FeSe alloys than the coherent-potential
approximation.

\section{Conclusions}

In conclusion, we have studied the electronic structure of Fe$_{1+\delta}$Se,
FeSe$_{1-x}$, FeSe$_{1-y}$Te$_{y}$ , as a function of $\delta$,
$x$ and $y$ as well as FeSe$_{0.9}$S$_{0.1}$, Fe$_{0.9}$Co$_{0.1}$Se,
Fe$_{0.9}$Ni$_{0.1}$Se and Fe$_{0.91}$Cu$_{0.1}$Se alloys. our
results show that (i) a small amount of excess Fe substantially disorders
the Fe-derived bands near $E_{F}$ while Se-deficiency affects mainly
the Se-derived bands away from $E_{F}$, (ii) the substitution of
S and Te enhances the possibility of FS nesting in spite of disordering
the Se-derived bands, specially in FeSe$_{0.5}$Te$_{0.5}$ alloy
, (iii) the substitution of Co, Ni or Cu disorders and pushes down
the Fe-derived bands, thereby destroying the possibility of FS nesting.
We also find that the coherent-potential approximation is more reliable
than the rigid-band, virtual-crystal or supercell approximations for
describing substitutional disorder in FeSe alloys. Thus, within the
framework of spin-fluctuation theories, our results provide a consistent
basis for understanding  the superconducting properties of FeSe alloys.

\ack{}{}

We like to thank A. V. Mahajan for helpful discussion.

\section*{References}{}

\bibliographystyle{unsrt}

\end{document}